\newif\ifpictures
\picturestrue

\documentclass[12pt]{amsart}
\usepackage{amssymb}
\usepackage[dvips]{graphicx}

\ifpictures
\usepackage{psfrag}
\fi

\numberwithin{equation}{section}
\newtheorem{thm}{Theorem}
\newtheorem{prop}[thm]{Proposition}
\newtheorem{lemma}[thm]{Lemma}

\theoremstyle{definition}

\newtheorem{example}[thm]{Example}
\newtheorem{remark1}[thm]{Remark}
\newtheorem{openproblem1}[thm]{Open problem}
\newtheorem{definition}[thm]{Definition}

\newenvironment{ex}{\begin{example}\rm}{\end{example}}
\newenvironment{rem}{\begin{remark1}\rm}{\end{remark1}}

\numberwithin{thm}{section}

\newcounter{FNC}[page]
\def\newfootnote#1{{\addtocounter{FNC}{2}$^\fnsymbol{FNC}$%
     \let\thefootnote\relax\footnotetext{$^\fnsymbol{FNC}$#1}}}

\newcommand{\N}{\mathbb{N}}

\newcommand{\R}{\mathbb{R}}

\DeclareMathOperator{\LP}{LP}

\title[Enumerating the Nash equilibria of rank 1-games]{Enumerating the Nash equilibria of
  rank 1-games}

\author{Thorsten Theobald}

\address{J.W.\ Goethe-Universit\"at, FB 12 -- Institut f\"ur Mathematik,
Postfach 11 19 32, D--60054 Frankfurt am Main, Germany}

\email{theobald@math.uni-frankfurt.de}

\begin{document}

\begin{abstract}
A bimatrix game $(A,B)$ is called a game of rank $k$ if the rank of 
the matrix
$A+B$ is at most $k$. We consider the problem of enumerating the Nash 
equilibria in (non-degenerate) games of rank~1.
In particular, we show that even for games of rank~1 
not all equilibria can be reached by a 
Lemke-Howson path and present a parametric simplex-type algorithm
for enumerating all Nash equilibria of a non-degenerate game of rank~1.
\end{abstract}

\maketitle




\section{Introduction}

Bimatrix games belong to the fundamental concepts of game theory.
A bimatrix game is given by two payoff matrices $(A,B)$, and by Nash's
results any bimatrix game has at least one equilibrium \cite{nash-pams-50,
nash-annals-51}. The problem of computing a Nash equilibrium
(named by Papadimitriou in 2001
to be the most concrete open question on the boundary of the complexity
class $\mathcal{P}$
\cite{papadimitriou-2001}) has received much attention in the last years.
Chen and Deng recently showed that
the problem is $\mathcal{PPAD}$-complete \cite{chen-deng-2005}
and (together with Teng \cite{cdt-2006})
that the problem of computing a $1/n^{\Theta(1)}$-approximate
Nash equilibrium remains $\mathcal{PPAD}$-complete. Thus it is unlikely that
an efficient algorithm exists.

The corresponding \emph{enumeration problem} asks to enumerate all
equilibria (in the finite case) or all the extreme equilibria (in the
degenerate case where an infinite number of Nash equilibria is possible).
The enumeration problem is similar to (but more difficult than) enumerating the
vertices of a polyhedron given as the intersection of half-spaces.
In the latter vertex enumeration problem, the Upper Bound Theorem
gives a tight estimate for the maximal number of vertices, but the
analogous problem of determining the maximum number of Nash equilibria
of a bimatrix game is an open problem (see \cite{stengel-99,stengel-handbook-2002}).

For the special case of zero-sum games, the set of Nash equilibria
defines a polyhedral set in the strategy space, and the problem of
computing the set of all Nash equilibria is equivalent to linear
programming (see \cite[Ch.~13.2]{dantzig-b51}). Hence, for a non-degenerate
zero-sum game, the set of Nash equilibria consists of a single point,
and thus the enumeration problem becomes trivial.

Recently, Kannan and Theobald \cite{kannan-theobald-soda-2007} have introduced a hierarchy of bimatrix games
in which the matrix $A+B$ is restricted to be of rank at 
most $k$, for some fixed constant~$k$.
For any fixed $k \ge 1$, this class strictly generalizes the
class of zero sum-games. Of course, the case $k=1$ is of particular
importance; it is the smallest extension of zero-sum games in the hierarchy.

In contrast to zero-sum games, non-degenerate rank $k$-games (for any fixed $k \ge 1$)
can have an arbitrarily large number of equilibria. In particular, this provides
a sharp separation between the class of rank~1-games and the class of zero-sum games.
From the computational viewpoint,
Nash equilibria can be efficiently approximated in games of fixed rank,
but the question of exact polynomial time computability is open even 
for games of rank~1.

In this paper, we consider the enumeration problem for games of rank~1.
Similar to the situation above, the rank condition provides additional
structure which can be exploited. An initial question is whether any
equilibrium can be reached by a Lemke-Howson path (as defined 
formally in Section~\ref{se:lemke-howson-paths}).
For arbitrary bimatrix games Aggarwal 
has shown that in general not all equilibria can be reached by a 
Lemke-Howson path \cite{aggarwal-73}. 
By providing an example of a rank 1-game for which not all
equilibria can be reached in this way, we strengthen Aggarwal's result
and thus answer our initial question in the negative.

As main contribution of the paper, we propose a parametric simplex-type 
algorithm for enumerating the Nash equilibria of rank 1-games.
This algorithm is based on the techniques of Konno and Kuno who 
have investigated linear multiplicative 
programs (\cite{konno-kuno-92}, see also \cite{ktt-97}).
Our problem can be seen as an enumeration problem of 
generalized linear multiplicative
programming. Moreover, the situation of games provides 
additional combinatorial structure which can be exploited.

The paper is structured as follows. In Section~\ref{se:prelim}, 
we introduce the basic
concepts of bimatrix games as well as of rank $k$-games 
and review existing work
on the enumeration of Nash equilibria.
In Section~\ref{se:lemke-howson-paths}, we show that not all equilibria can be
reached by a Lemke-Howson path in a rank 1-game.
Then, in Section~\ref{se:parametric}, we present the parametric
simplex-type algorithm for enumerating
the Nash equilibria of a non-degenerate rank~1-game; finally,
we explain how to modify the algorithm so as to cover
degenerate situations as well.

\section{Preliminaries\label{se:prelim}}

\subsection{Bimatrix games}

We consider an $m \times n$-bimatrix game with payoff matrices 
$A,B \in \R^{m \times n}$.
Let 
\[
  \mathcal{S}_1 = \big\{ x \in \R^m \, : \, \sum_{i=1}^m x_i = 1 \, , \: 
  x \ge 0 \big\} \;  \text{ ~~and~~ } \; 
  \mathcal{S}_2 = \big\{ y \in \R^n \, : \, \sum_{j=1}^n y_j = 1 \, , \:
  y \ge 0 \big\}
\]
be the sets of mixed strategies of the two players, and let
$\overline{\mathcal{S}}_1 = \{x \in \R^m \, : \, \sum_{i=1}^m x_i = 1 \}$
and 
$\overline{\mathcal{S}}_2 = \{y \in \R^n \, : \, \sum_{j=1}^n y_j = 1 \}$
denote the underlying affine subspaces. The first player (the row player)
plays $x \in  \mathcal{S}_1$ and the second player (the column player)
plays $y \in \mathcal{S}_2$. The payoffs for player~1 and player~2 are
$x^T A y$ and $x^T B y$, respectively.

Let $C^{(i)}$ denote the $i$-th row of a matrix $C$ (as a row vector), 
and let $C_{(j)}$ denote the $j$-th column of $C$ (as a column vector).
A pair of mixed strategies $(\overline{x}, \overline{y})$
is a \emph{Nash equilibrium} if
\begin{equation}
\label{eq:defnash}
  \overline{x}^T A \overline{y} \ \ge \ x^T A \overline{y}
  \quad \text{ and } \quad \overline{x}^T B \overline{y} \ \ge \ \overline{x}^T B y
\end{equation}
for all mixed strategies $x$, $y$.
Equivalently, $(\overline{x},\overline{y})$ is a Nash equilibrium 
if and only if
\begin{equation}
\label{eq:bestresponse1}
  \overline{x}^T A \overline{y} \ = \ 
  \max_{1 \le i \le m} A^{(i)} \overline{y}
  \quad \text{ and } \quad
  \overline{x}^T B \overline{y} \ = \ 
  \max_{1 \le j \le n} \overline{x}^T B_{(j)} \, .
\end{equation}

A bimatrix game is called \emph{non-degenerate}
if the number of the pure best responses of player~1 to a mixed
strategy $y$ of player~2 never exceeds the cardinality of the
support $\mathrm{supp~}y := \{j \, : \, y_j \neq 0\}$
and if the same holds true for the best pure responses of player~2 
(see~\cite{stengel-handbook-2002}). In the case of a non-degenerate
game the set of Nash equilibria consists of finitely many isolated points.
We remark that there exist various other definitions of degeneracy in the literature 
which are equivalent to that notion 
(see~\cite[Theorem~2.10]{stengel-handbook-2002}).

\subsection{Earlier work on enumeration of equilibria\label{se:earlierwork}}

The classical Lemke-Howson algorithm serves to find one Nash equilibrium
in a bimatrix game (\cite{lemke-howson-64}, 
see also \cite{stengel-handbook-2002}). We discuss this algorithm
and its (negative) relation to enumeration 
in more detail in Section~\ref{se:lemke-howson-paths}.

From the viewpoint of computational complexity,
the problem of counting the number of Nash equilibria in a
bimatrix game is $\# \mathcal{P}$-hard \cite{conitzer-sandholm-2003}.
Hence, the enumeration problem is $\# \mathcal{P}$-hard.

The general idea of existing approaches for the enumeration of 
Nash equilibria is to transform the
problem into a  problem of polyhedral computation.
For each game we define a pair of polyhedra in which each Nash equilibrium 
of the game corresponds to a \emph{complementary vertex pair}.

\begin{definition} \label{de:mangasarian}
For an $m \times n$-bimatrix game $(A,B)$, define the
polyhedra $P$ and $Q$ by
\begin{eqnarray}
  \: \quad P & = & \{(x, \pi_2) \in \R^m \times \R \, \, : \,
\underbrace{x \ge 0}_{\text{inequalities } 1, \ldots, m}, \;
  \underbrace{x^T B \le {\bf 1}^T \pi_2}_{\text{inequalities } m+1, \ldots, m+n}, \; {\bf 1}^T x = 1 \} \, \label{eq:p} \, , \\
  \: \quad Q & = & \{(y, \pi_1) \in \R^n \times \R \, \, : \,
  \underbrace{A y \le {\bf 1} \pi_1}_{\text{inequalities } 1, \ldots, m}
 , \; \underbrace{y \ge 0}_{\text{inequalities } m+1, \ldots, m+n} ,
 \; {\bf 1}^T y = 1 \} \label{eq:q} \, ,
\end{eqnarray}
where ${\bf 1}$ is the all-1-vector.
\end{definition}

A pair of mixed strategies 
$(\overline{x},\overline{y}) \in \mathcal{S}_1 \times \mathcal{S}_2$ 
is a Nash equilibrium if and only if there exist
$\pi_1, \pi_2 \in \R$ such that
$(\overline{x},\pi_2) \in P$,
$(\overline{y},\pi_1) \in Q$ and for all $i \in \{1, \ldots, m+n\}$,
the $i$-th inequality of $P$ or $Q$ is binding.
Here, $\pi_1$ and $\pi_2$ represent the payoffs of player~1 and player~2, respectively.
For $i \in \{1, \ldots, m\}$ we call the inequality $x_i \ge 0$ the
\emph{$i$-th nonnegativity inequality} of $P$, and for $j \in \{1, \ldots, n\}$
we call the inequality $\overline{x}^T B_{(j)} \le \pi_2$ the 
\emph{$j$-th best response inequality} of $P$. 
And analogously for $Q$.

\begin{ex} (Taken from \cite{theobald-semesterberichte-2005}).
The bimatrix game with payoff matrices
\[
  A \ = \begin{pmatrix}
    2 & 1 & 5 \\
    3 & 0 & 4
  \end{pmatrix} \, , \qquad
  B \ = \begin{pmatrix}
    7 & 8 & 1 \\
    2 & 1 & 6
  \end{pmatrix}
\]
has 3 Nash equilibria:
\[
  \left( \left(1,0\right)^T,\left(0,1,0\right)^T \right), \quad
  \left( \left(\frac{1}{2},\frac{1}{2}\right)^T,
  \left(\frac{1}{2},\frac{1}{2},0 \right)^T \right), \quad
  \left( \left(\frac{2}{5},\frac{3}{5}\right)^T,
  \left(\frac{1}{2},0,\frac{1}{2}\right)^T\right) \, .
\]

\ifpictures
\begin{figure}[ht]
\[
  \begin{array}{c@{\hspace*{0.5cm}}c}
    \includegraphics[scale=1]{pictures/picparametric.5} \\ [-5.3cm]
    & \includegraphics[scale=1]{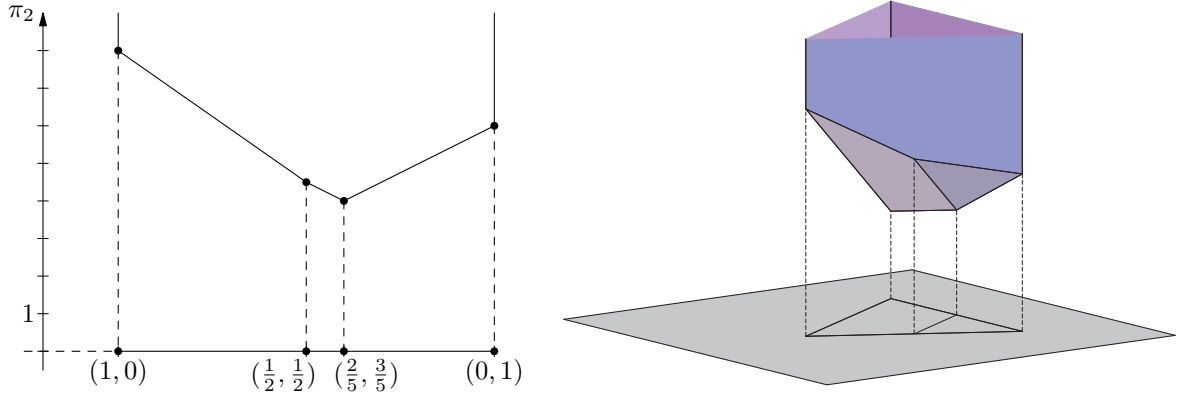}
  \end{array}
\]
\caption{In the example,
  $P$ is a two-dimensional polyhedron embedded in $\overline{\mathcal{S}}_1$,
  and $Q$ is a three-dimensional polyhedron embedded in $\overline{\mathcal{S}}_2$.
  The vertical direction corresponds to the variables $\pi_2$ and $\pi_1$, 
  respectively.}
\label{fi:lift1}
\end{figure}
\fi
The polyhedra $P$ and $Q$ are shown in Figure~\ref{fi:lift1}.
\end{ex}

The situation becomes more involved when one also wants to handle degenerate
games. For an arbitrary bimatrix game,
the set of all Nash equilibria is in general a non-convex subset, but can be represented
as the union of a finite number of polytopes (called \emph{maximal Nash subsets})
in the strategy space
(see \cite{millham-74}). Hence, in the degenerate situation, the task is
to enumerate all vertices
of every maximal Nash subset of a bimatrix game. Every equilibrium point
is a convex combination of some of these \emph{extreme equilibria}
(cf.\ \cite{jansen-81, winkels-79}).

These concepts can be used to provide algorithms for enumerating Nash equilibria 
of bimatrix games. The earliest ones can be found in
Vorob'ev \cite{vorobev-58} and Mangasarian \cite{mangasarian-64};
for later developments see Mukhamediev \cite{mukhamediev-78}, Winkels \cite{winkels-79},
and 
Audet, Hansen, Jaumard, and Sauvard \cite{ahjs-01}.
The latter paper also contains a detailed historical account on these algorithms.

\subsection{Games of fixed rank}
For a given constant $k \ge 0$, a bimatrix game is called 
a \emph{game of rank} $k$ if the matrix $A+B$ has rank at most $k$.

Kannan and Theobald have shown that for any fixed $k \ge 1$,
the number of Nash equilibria in a non-degenerate game of rank $k$
cannot be bounded by a function in terms of $k$
\cite{kannan-theobald-soda-2007}. In particular, the case $k=1$
stands in sharp contrast to the 
case $k=0$ of zero-sum games; there, the number of Nash equilibria
in the finite situation is always 1. For the case $d:=m=n$,
the best known lower bound for the maximal number of Nash equilibria
of rank~1-games is linear in $d$:

\begin{prop} \label{th:linearbound}
For any $d \in \N$ there exists a non-degenerate $d \times d$-game of
rank~1 with at least $2d-1$ Nash equilibria.
\end{prop}

A construction achieving this number is given by the
$d \times d$-game $(A,B)$ with
\begin{equation}
\label{eq:rank1constr}
  a_{ij} \ = \ 2ij - i^2 + j^2 \, , \qquad
  b_{ij} \ = \ 2ij + i^2 - j^2 
\end{equation}
(see \cite{kannan-theobald-soda-2007}).
Since $A + B = (4ij)_{i,j}$, the matrix $A+B$ is of rank~1.

It is not known whether in games of fixed rank a Nash equilibrium
can be computed in polynomial time. In \cite{kannan-theobald-soda-2007},
the following approximation result was shown. Here,
a pair $(x,y)$ of mixed strategies 
is called an \emph{$\varepsilon$-approximate equilibrium} if 
\[
  \ell(x,y) \ \le \ \varepsilon |A+B| \, , 
\]
where $\ell(x,y)$ denotes the sum of the losses of the players,
\[
  \ell(x,y) \ = \ \max_i A^{(i)} y + \max_j x^T B_{(j)} - x^T (A+B) y \, ,
\]
and $|\cdot|$ denotes the maximum absolute value of the
entries of a matrix.

\begin{prop} \label{th:approxlowranknash}
Let $k$ be a fixed constant and $\varepsilon > 0$. 
In a game of rank $k$, an $\varepsilon$-approximate 
Nash equilibrium can be found in time
$\mathrm{poly}(\mathcal{L},1/\varepsilon)$,
where $\mathcal{L}$ is the bit length of the input.
\end{prop}

There are several operations on a bimatrix game
$(A,B)$ which do not change the set of Nash equilibria:
\begin{enumerate}
\item adding multiples of the all-1-vector to a given 
  column of $A$ or a given row of $B \, ;$
\item positive scaling of a given column of $A$ or a 
  given row of $B \, .$
\end{enumerate} 

Since these operations can change the rank of the
game, the following useful consequence is obtained.

\begin{lemma} \label{le:rankreduction}
Let $(A,B)$ be a $d \times d$-game of rank $d$. Then there exists a game of
rank $d-1$ with the same set of Nash equilibria.
\end{lemma}

\begin{proof}
Assume without loss of generality that 
$d \ge 2$ and that $C:=A+B$ is of rank exactly~$d$.
Then there exists a column $j$ of $C$ which is 
not a multiple of the all-1-vector. Denote by $v$ 
the column vector obtained from the entries of the $j$-th 
column of $C$.
Since the affine line
$v + \R (1, \ldots, 1)^T$ intersects the $(d-1)$-dimensional linear 
subspace defined by the $d-1$ other columns, there exists
some $\lambda \in \R$ such that adding $\lambda(1, \ldots,1)^T$
to the $j$-th column of $C$ yields a matrix of rank at most $d-1$. 
Thus,
adding $\lambda(1, \ldots, 1)^T$ to
the $j$-th column of $A$ turns the game $(A,B$) into a game of
rank $d-1$.
\end{proof}

\section{Lemke-Howson paths\label{se:lemke-howson-paths}}

In this section, we recall the classical Lemke-Howson algorithm for
finding a Nash equilibrium in a bimatrix game and then show that
not even for games of rank~1
all equilibria can be reached by a Lemke-Howson path.

The Lemke-Howson algorithm is a simplex-type algorithm which can 
be combinatorially described in terms of a graph. In order to define
this graph, we start from the polyhedral description of the bimatrix 
game in terms of $P$ and $Q$. Consider a pair of strategies 
$(x,y) \in \mathcal{S}_1 \times \mathcal{S}_2$, and let $\pi_1$ and
$\pi_2$ be the resulting payoffs.
We label each of the strategies $x$ and $y$ by the indices of the
inequalities in~\eqref{eq:p} and~\eqref{eq:q} that are binding.
For a non-degenerate $m \times n$-game, only the vertices of $P$ have $m$ labels and the vertices of $Q$
have $n$ labels, and there do not exist points in $P$ and $Q$
with more than $m$ or $n$ labels, respectively (see \cite[Theorem~2.7]{stengel-handbook-2002}).

We define the graphs $G_1 = (V_1,E_1)$  and $G_2 = (V_2, E_2)$ as follows.
The vertex set $V_1$ of $G_1$ consists of the vertices of $P$, with an 
additional vertex $0 \in \R^m$ that has all labels in 
the set $\{1, \ldots, m\}$. Two vertices $x$ and $x'$ are connected by
an edge if they differ in exactly one label, i.e., if they have
$m-1$ labels in common.
Similarly, let $G_2$ be the graph whose vertex set $V_2$ consists of
the vertices of $Q$, with an additional vertex $0 \in \R^n$ having
all labels in $\{m+1, \ldots, m+n\}$. Two vertices in $G_2$ are
connected if they have $n-1$ labels in common.

The product graph $G_1 \times G_2$ of $G_1$ and $G_2$ is defined
by the vertex set $V_1 \times V_2$, and the edges are given
by $\{x\} \times \{y,y'\}$ for vertices $x$ of $G_1$ and edges
$\{y,y'\}$ of $G_2$, or by $\{x,x'\} \times \{y\}$ for edges
$\{x,x'\}$ of $G_1$ and vertices $y$ of $G_2$.

From a combinatorial viewpoint, the Lemke-Howson algorithm can now be
described as follows.
Fix an $r \in \{1, \ldots, m+n\}$.
A vertex $(x,y)$ of $G_1 \times G_2$ is called
\emph{$r$-almost completely labeled} if the union of the labels
is the set $\{1, \ldots, m+n\} \setminus \{r\}$. Since two adjacent
vertices $x$ and $x'$ in $G_1$ have $m-1$ common labels, the edge
$\{x,x'\} \times \{y\}$ of $G_1 \times G_2$ is also $r$-almost completely
labeled if $y$ has the remaining $n$ labels except $r$.
And similarly for edges $\{x\} \times \{y,y'\}$ of $G_1 \times G_2$.
The Lemke-Howson algorithm starts from the artificial equilibrium $(0,0)$
which has all labels and then follows the unique 
path where the label $r$ is missing.
After finitely many steps, it reaches a Nash equilibrium of the game.

For different choices of $r$ it is possible that we reach different
Nash equilibria. This led to the early question in the algorithmic study of
games on whether any equilibrium can be reached by a Lemke-Howson path, i.e., by
some choice of $r$.
For general bimatrix games it is known that the set of Lemke-Howson paths
does not enumerate all Nash equilibria (Aggarwal \cite{aggarwal-73}; 
see also \cite{stengel-handbook-2002} and the references therein).

Since games of rank~1 are a very special case of general bimatrix games,
the question arises whether for a game of rank~1 all equilibria can be
reached by a Lemke-Howson path.
Here, we strengthen the (un-)reachability result by showing that even for
games of rank~1 not all equilibria can be reached. Namely, we consider
the rank 1-game
\[
  A \ = \ \left( \begin{array}{rr}
   -28 & -18 \\
    -8 & -23
  \end{array} \right) \, , \qquad
  B \ = \ \left( \begin{array}{rr}
    10 & 30 \\
    20 & 15
  \end{array} \right) \, ,
\]
which is a variation of Aggarwal's example resulting 
from Lemma~\ref{le:rankreduction}.
The polyhedra $P$ and $Q$ are shown in Figure~\ref{fi:unreachability}.

\ifpictures
\begin{figure}[ht]
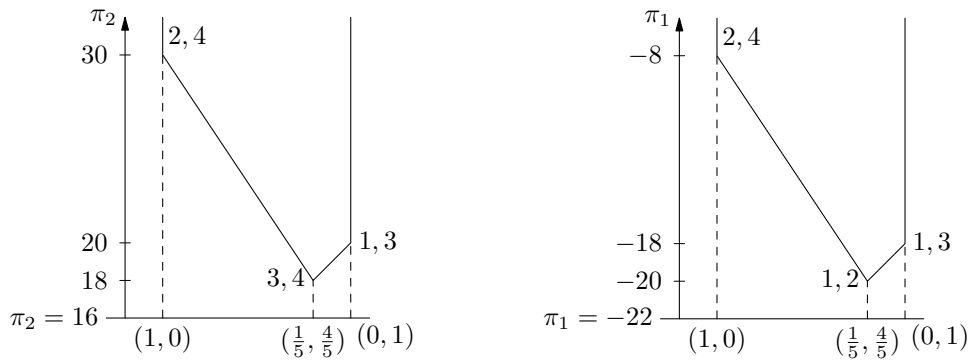

\[
  \begin{array}{c@{\qquad \qquad}c@{\qquad}c}
    \includegraphics[scale=1]{pictures/picparametric.6} &
    \includegraphics[scale=1]{pictures/picparametric.7}
  \end{array}
\]

\caption{The polyhedra $P$ and $Q$ and the labels of the vertices.}
\label{fi:unreachability}
\end{figure}
\fi

The game has three Nash equilibria:
\[
\begin{array}{rlll}
  & \left( (1,0)^T , (0, 1)^T \right) \, , & \text{ with payoffs } (-18,30) \, ; \\ [2ex]
  & \left( (0,1)^T , (1,0)^T \right) \, , & \text{ with payoffs } (-8,20) \, ; \\ [2ex]
  \text{and} & \left( (\frac{1}{5},\frac{4}{5})^T , (\frac{1}{5}, \frac{4}{5})^T \right) \, , & \text{ with payoffs } (-20,18) \, . 
\end{array}
\]

With regard to the Lemke-Howson graphs, the locally maximal peaks of the lower hull of $P$
are connected with the $0$-vertex (which has labels $1,2$) and the maximal peaks
in the lower hull of $Q$ are connected with the $0$-vertex (which has labels $3,4$).
Considering all the four possible values of $r$, only the first two
equilibria can be obtained via Lemke-Howson paths, and the third equilibrium 
cannot be obtained. E.g., for the initial missing label $r=1$ we obtain the
following path with labels. Here, the two components correspond to the graphs
$G_1$ and $G_2$.
\[
  \left( \begin{array}{cc} 1,2 \\ 3,4 \end{array} \right) \, \rightarrow \, 
  \left( \begin{array}{cc} 2,4 \\ 3,4 \end{array} \right) \, \rightarrow \, 
  \left( \begin{array}{cc} 2,4 \\ 1,3 \end{array} \right) \, ,
\]
where the last vertex pair gives the Nash equilibrium $((1,0)^T, (0,1)^T)$.
Hence we can conclude:

\begin{thm}
There exist games of rank~1 for which not all equilibria can be
reached by a Lemke-Howson path.
\end{thm}

Let $G'$ be the graph with vertex set $V_1 \times V_2$ whose edge
set is the union (over all $r$) of $r$-almost completely labeled
edges. In a paper of Shapley \cite{shapley-74}, 
the following example attributed to Wilson is
given which shows that for games of arbitrary rank the graph $G'$
can even be disconnected. Let the $3 \times 3$-game $(A,B)$
be defined by
\[
  A \ = \ \left( \begin{array}{ccc}
    0 & 3 & 0 \\
    2 & 2 & 0 \\
    3 & 0 & 1
  \end{array} \right) \, , \qquad
  B \ = \left( \begin{array}{ccc}
    0 & 2 & 3 \\
    3 & 2 & 0 \\
    0 & 0 & 1
  \end{array} \right) \, .
\]

\ifpictures
\begin{figure}[ht]
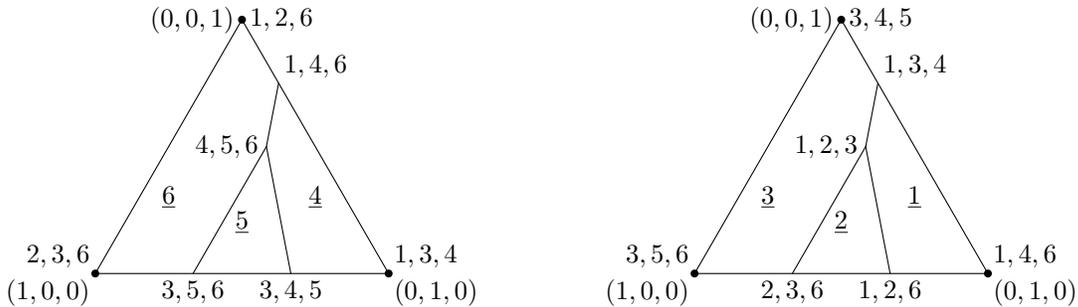

\[
  \begin{array}{c@{\qquad \qquad}c@{\qquad}c}
    \includegraphics[scale=1]{pictures/picparametric.8} &
    \includegraphics[scale=1]{pictures/picparametric.9}
  \end{array}
\]

\caption{The projections of the polyhedra $P$ and $Q$ in Wilson's 
  examples. The picture contains the labels of the points with
  three labels. The underlined numbers show the
  best responses of the regions.}
\label{fi:wilson}
\end{figure}
\fi

In the graph $G'$, the mixed equilibrium of $(A,B)$ cannot
be reached from the artificial equilibrium. Hence,
even modifications of
the Lemke-Howson-algorithm which are
allowed to change the index of the missing label within
the graph traversal cannot enumerate all Nash equilibria.

Applying the rank reduction Lemma~\ref{le:rankreduction} on
Wilson's example immediately implies that for 
$3 \times 3$-games of rank~2, the graph $G'$ can be disconnected.
It would be interesting to know if this graph can also be disconnected
for the case of rank 1-games. While from the principle viewpoint
this question is decidable, currently we do not know the answer.

\section{A parametric algorithm for enumerating all Nash equilibria of
  rank~1-games\label{se:parametric}}

In this section, we present a parametric simplex-type enumeration algorithm
for games of rank~1. For simplicity we concentrate on the situation where
the game is non-degenerate and the polytopes $P$ and $Q$ are in 
general position. In Section~\ref{se:degenerate} we explain how to modify the algorithm
to cover degenerate situations as well.
An example is presented in 
Section~\ref{se:example}.
 
\subsection{Non-degenerate situations}

We consider the following well-known characterization of a Nash equilibrium
in terms of a quadratic program \cite{mangasarian-stone-64}.
For any $(x,\pi_2) \in P$ and $(y,\pi_1) \in Q$, we have
\[
  x^T (A+B) y \ \le \ \max_{1 \le i \le m} A^{(i)} x + \max_{1 \le j \le n} x^T B_{(j)} \ \le \ \pi_1 + \pi_2
\]
with equality if and only $(x,y)$ is a Nash equilibrium with payoffs $\pi_1$ and $\pi_2$.
Consider the quadratic program
\begin{equation}
\label{eq:qp}
  \begin{array}{l@{\quad}rcl}
  \mathrm{(QP:)} & \multicolumn{3}{c}{\max~ x^T (A+B) y - \pi_1 - \pi_2} \\
  & (x,\pi_2) & \in & P \, , \\
  & (y,\pi_1) & \in & Q \, .
  \end{array}
\end{equation}
Hence, we obtain the following characterization of the Nash equilibria in
terms of the quadratic program.

\begin{lemma}
\label{le:qp1}
1) For any feasible solution of $\mathrm{QP}$, the objective value is nonpositive.
\smallskip

\noindent
2) A pair $(x^*,y^*) \in \mathcal{S}_1 \times \mathcal{S}_2$ 
is a Nash equilibrium of the bimatrix game $(A,B)$ if and only if there exist
$\pi_1^*, \pi_2^* \in \R$ such that $(x^*,y^*,\pi_1^*,\pi_2^*)$ is a feasible
solution of $\mathrm{QP}$ with objective value~0.
\end{lemma}

Note that the inequalities describing the feasible set are linear.
Moreover, the objective function
only depends on the sum $A+B$ rather than on $A$ or $B$ individually.

If the game is of rank~1, we write $A+B$ in the form
$A + B = b \cdot c^T$ with $b \in \R^m$, $c \in \R^n$.
That is, after a linear variable transformation we want to
enumerate the global optima of a function of the form $f+g$ where $f$ is the
product of two variables and $g$ is a linear function.
The problem of optimizing a product of two variables 
over a polyhedron is called a \emph{linear multiplicative program}
(\cite{konno-kuno-92}, see also \cite{ktt-97}).
In these references, parametric primal-dual simplex-type algorithms
were presented to find the optimal value. 
Based on these techniques, we now present
an algorithm which enumerates all equilibria of a rank 1-game.
Since we are starting from a game, we will see below that
the relevant bases have a special combinatorial structure,
where $m$ elements of the basis refer to the
polyhedron~$P$ and $n-1$ elements refer to the polyhedron $Q$.

By setting $\xi := c^T y$, we can write the quadratic program as
\begin{equation}
\label{eq:qpxi}
  \begin{array}{rcl}
  \multicolumn{3}{c}{\max~ (x^T b) \xi - \pi_1 - \pi_2} \\
  (x,\pi_2) & \in & P \, , \\
  (y,\pi_1) & \in & Q \, , \\
  c^T y & = & \xi \, . \\
  \end{array}
\end{equation}
We consider $\xi$ as a parameter to the optimization problem. For a given
value to $\xi$, the problem becomes a linear program which we call $\text{LP}(\xi)$.
Geometrically, for fixed $\xi$ we are slicing the feasible set polyhedron 
by a hyperplane $c^T y = \xi$ on which the strategy of the second player
satisfies a particular linear condition.

\begin{rem}
In the special case where $c$ is a multiple of the all-1-vector, the
hyperplane defined by $c^T y = \xi$ is parallel to the hyperplane 
defined by ${\bf 1}^T y = 1$ (which is part of the definition of $Q$).
In that situation, all columns of $A+B$ are identical, and games of 
this type are called \emph{row-constant games} (see \cite{isaacson-millham-80}).
We will come back to that special case below.
\end{rem}

The feasible set of~\eqref{eq:qp} is unbounded. However,
since $y \in \mathcal{S}_2$ the range $[\xi_{\min},\xi_{\max}]$ of $\xi$ is bounded,
namely
\begin{equation}
\label{eq:range}
  [\xi_{\min}, \xi_{\max}] \ = \ 
  [ \min_{y \in \mathcal{S}_2} c^T y , \max_{y \in \mathcal{S}_2} c^T y] 
  \ = \ [\min_{1 \le j \le n} c_j,
         \max_{1 \le j \le n} c_j ] \, .
\end{equation}
Even for fixed $\xi$, the feasible region of $\mathrm{LP}(\xi)$
can be unbounded. But by Lemma~\ref{le:qp1},
the objective value of QP is bounded
from above and hence also $\mathrm{LP}(\xi)$ is bounded from
above.

Let $\mathcal{I} := \{1, \ldots, m, m+1, \ldots, m+n\}$ be
the index set of the inequalities of $P$, and 
$\mathcal{J} := \{1, \ldots, m, m+1, \ldots, m+n\}$ be the index set of the
inequalities of $Q$.

We consider a fixed parameter value $\xi$. If the resulting 
$\text{LP}(\xi)$ is in general position,
then the optimal point $v$ of $\text{LP}(\xi)$ is unique and
$v$ is a vertex of the feasible set polyhedron of 
$\mathrm{LP}(\xi)$.
If $\xi$ is a sufficiently generic 
value (we will come back to this aspect below) then 
$v$ can be uniquely described in terms of a basis, 
i.e., by a choice $I \subset \mathcal{I}$ of cardinality $m$ and 
$J \subset \mathcal{J}$ of cardinality $n-1$. 

For a given $\xi$, let $(I,J)$ denote an optimal basis
for the linear program $\mathrm{LP}(\xi)$ depending on $\xi$.
The idea of the enumeration algorithm is to keep track on
the values of $\xi$ where the optimal basis changes. 
This yields the enumeration algorithm in 
Figure~\ref{fi:enumerationalgo}.

\begin{figure}[htb]
\noindent
\rule{\textwidth}{0.1pt}

\begin{small}
\begin{sf}
\begin{flushleft}
Set $\xi = \xi_{\min}$; \\
While $\xi \le \xi_{\max}$ do \\
\qquad Compute the optimal point $z := z(\xi)$ of $\mathrm{LP}(\xi)$; \\
\qquad If $z(\xi)$ has objective value 0 then \\
\qquad \qquad Let $x,y$ be the strategies played in $z(\xi)$; \\
\qquad \qquad Output ``Nash equilibrium:'', $x,y \, ;$ \\
\qquad  Compute the smallest $\xi' > \xi$ where the optimal basis changes 
  and the new optimal basis;

\end{flushleft}
\end{sf}
\end{small}

\noindent
\rule{\textwidth}{0.1pt}

\caption{Enumeration algorithm}
\label{fi:enumerationalgo}
\end{figure}

In order to explain how the update steps can be performed efficiently,
we analyze the set of
the values of the parameter $\xi$ which yield the same optimal basis.
For this, we consider the dual program of the parametric LP~\eqref{eq:qpxi}.

Let $u_i$ be the dual variable corresponding to the $i$-th inequality
of $P$, $1 \le i \le m+n$, 
and $u_{m+n+j}$ be the dual variable corresponding to the $j$-th
inequality of $Q$, $1 \le j \le m+n$. Further let $u_{2(m+n)+1}$ and $u_{2(m+n)+2}$ be the 
dual variables of the equations ${\bf 1}^T x =1$ and
${\bf 1}^T y= 1$, and $u_{2(m+n)+3}$ be the
dual variable of the equation $c^T y = \xi$. The dual variables $u_1, \ldots, u_{2(m+n)}$
are sign-restricted while the dual variables $u_{2(m+n)+1}, \ldots, u_{2(m+n)+3}$ are not.

We set $z = (x,y,\pi_1,\pi_2)$ and the constants
$K = 2(m+n)$ and $N = m+n+2$. Let
$M_1 \in \R^{K \times N}$ and $e_1 \in \R^K$ be defined by
\[
  M_1 \ = \ \left( \begin{array}{c|c|c|c}
    -I & & & \\
    B^T & & & - {\bf 1} \\
    & A & - {\bf 1} & \\
    & -I & &
  \end{array} \right) \, , \quad
  e_1 = 0 \, .
\]
Further, let
$M_2 \in \R^{3 \times N}$ and $e_2 \in \R^3$ be defined
by
\[
  M_2 \ = \ \left( \begin{array}{c|c|c|c}
    {\bf 1}^T & & \hspace*{2ex} & \hspace*{2ex} \\
    & {\bf 1}^T & & \\
    & c^T & & \\
  \end{array} \right) \, , \quad
  e_2 = \left( \begin{array}{c}
    1 \\ 1 \\ \xi
  \end{array} \right) \, .
\]
Then the feasible region of the linear program $\text{LP}(\xi)$ 
is given by $M_1 z \le e_1$, $M_2 z = e_2$. 
For fixed $\xi$, the dual of $\text{LP}(\xi)$ is
\[
  \begin{array}{rcl}
  \multicolumn{3}{l}{\min~ (e_1^T,e_2^T) u} \\
  (M_1^T | M_2^T) u & = & (b_1 \xi, \ldots, b_m \xi, 
  \underbrace{0, \ldots, 0}_{n \text{ times}}, -1, -1)^T \, , \\
  u_1, \ldots, u_K & \ge & 0 \, , 
  \end{array} 
\]
where $u = (u_1, \ldots, u_{K+3})$ is the vector of dual variables.

By the complementary slackness conditions for linear programming, 
the optimal solution of the dual program satisfies 
\begin{equation}
\label{eq:complementary}
  u_i=0 \text{ for all } i \not\in I \quad \text{ and } \quad u_{m+n+j} = 0 \text{ for all } j \not\in J \, .
\end{equation}

A basis $(I,J)$ of $\LP(\xi)$ is also an optimal basis for all those programs $\LP(\xi')$ 
for which the point described by $(I,J)$ is feasible and for
which there exists a feasible solution to the dual program satisfying
the complementarity condition~\eqref{eq:complementary}.

For the special case of a zero-sum game, the parametric formulation
degenerates to the well-known 
pair of dual linear programs associated with the game. Namely, if 
$b=c=0$ then $\xi_{\min} = \xi_{\max} = 0$ and the matrix $M_1$ becomes
\[
  M_1 \ = \ \left( \begin{array}{c|c|c|c}
    -I & & & \\
    (-A)^T & & & - {\bf 1} \\
    & A & - {\bf 1} & \\
    & -I & &
  \end{array} \right) \, .
\]
For $\xi = 0$, the dual then becomes
\[ \begin{array}{lrcl}
  & \multicolumn{3}{l}{\min u_{K+1} + u_{K+2}} \\
  \quad \text{s.t.} & \left( \begin{array}{c|c|c|c|c|c}
    -I & -A & & & {\bf 1} & \\
    & & A^T & -I & & {\bf 1} \\
    & & - {\bf 1}^T & & & \\
    & - {\bf 1}^T & & & & 
  \end{array} \right) u & = & 
  \left( \begin{array}{c}
    0 \\ 0 \\ -1 \\ -1
  \end{array} \right) \, , \\
  & u_1, \ldots, u_{K} & \ge & 0 \, .
  \end{array} 
\]

Hence we can conclude:
\begin{lemma}
If the game is a zero-sum game (i.e., $A+B = 0$) 
then the parametric problem~\eqref{eq:qpxi} is only feasible
for $\xi = 0$. In this case, the constraints of the 
dual LP coincide with the constraints of the primal program 
under the identifications
$x_i = u_{m+n+i}$, $1 \le i \le m$,
$y_j = u_{m+j}$, $1 \le j \le n$, 
$\pi_1 = u_{2(m+n)+2}$,
$\pi_2 = u_{2(m+n)+1}$,
and considering $u_1, \ldots, u_{m}$ and 
$u_{2m+n+1}, \ldots, u_K$ as slack variables; the objective
functions are additive inverses under these identifications.
\end{lemma}

Similarly, for row-constant games the range of $\xi$-values consists of a single
point, and the equilibria of these 
games can be phrased as linear programs. Namely, if there
exist constants $u_1, \ldots, u_m \in \R$ with
\[
  a_{ij} \ + \ b_{ij} \ = \ u_i \quad \text{for $1 \le i \le m, \; 1 \le j \le n$} 
\]
then the zero-sum
game $(A',B')$ defined by $b'_{ij} \ = \ b_{ij} - u_i$
has the same Nash equilibria as $(A,B)$.

From now on, let the game $(A,B)$ neither be a zero-sum game nor a 
row-constant game.
Let $(I,J)$ be an optimal basis of $\mathrm{LP}(\xi)$ for
some parameter value $\xi$.
Let $\mathcal{B} = I \cup J$, and for a matrix $A$ let $A_{\mathcal{B}}$ be the
submatrix of $A$ with rows in $\mathcal{B}$.
By our assumption, the system of linear equations
\begin{eqnarray}
  (M_1)_{\mathcal{B}} z & = & (e_1)_{\mathcal{B}} \, , \label{eq:lineq1} \\
  M_2 z & = & e_2 \label{eq:lineq2}
\end{eqnarray}
has a unique solution. Let $z(\xi)$ be the solution point
of this system.
In order to check whether there exists a dual solution satisfying
the complementary slackness conditions,
set $\mathcal{B}' = \mathcal{B} \cup \{K+1,K+2,K+3\}$. Compute
\[
  ((M_1)_{\mathcal{B'}}^T|M_2^T)^{-1} (b_1 \xi, \ldots, b_m \xi, 
   0, \ldots, 0, -1, -1)^T
\]
and set
all components of $u$ 
indexed by $\{1, \ldots, 2K \} \setminus \mathcal{B}$ 
to zero to obtain a vector $u=u(\xi)$ with
$(M_1^T | M_2^T) u(\xi) = (b_1 \xi, \ldots, b_m \xi, 0, \ldots, 0, -1, -1)^T$
and which satisfies the complementarity conditions.

\begin{lemma}
\label{le:optimality1}
Let $z(\xi)$ be defined by solving~\eqref{eq:lineq1} and~\eqref{eq:lineq2} for $z$ and $u(\xi)$ be as described before.
The set of $\xi$ such that $\mathcal{B}$ is an optimal basis 
of $\mathrm{LP}(\xi)$ is given by the two conditions
\begin{eqnarray}
  M_1 z(\xi) & \le & e_1 \, , \label{eq:optimality1} \\
  u_1(\xi), \ldots, u_K(\xi) & \ge & 0 \, . \label{eq:optimality2}
\end{eqnarray}

\end{lemma}

\begin{proof}
The first condition is satisfied if and only if $z(\xi)$ is feasible.
For a vertex $z(\xi)$ of the feasible set polyhedron
the second condition is satisfied if and only if
$z(\xi)$ is optimal for $\text{LP}(\xi)$.
\end{proof}

For a variable $\xi$, both conditions in~\eqref{eq:optimality1}
and~\eqref{eq:optimality2}
are linear conditions in $\xi$. Hence, the range of $\xi$-values in which
both conditions are satisfied defines an interval.
Let $[\alpha_1, \alpha_2]$ and
$[\beta_1,\beta_2]$ be the intervals defined by~\eqref{eq:optimality1}
and~\eqref{eq:optimality2}, respectively.
Then $[\xi_1,\xi_2] := 
[\max\{ \alpha_1,\beta_1\}, \min \{\alpha_2,\beta_2 \}]$
is the interval for $\xi$ in which both conditions are satisfied. 
Since $(I,J)$ is an optimal basis for some $\xi$,
the interval is nonempty.
We distinguish two cases:

\medskip

\noindent
{\bf Case $\xi_2 = \alpha_2$.} Then for the value 
$\xi = \xi_2$,
there are $m+n$ inequalities, indexed by $\mathcal{B} \cup \{j\}$ for
some $j$, which are binding in the primal program.
One of the inequalities $i$ of the current vector $z(\xi)$ becomes
violated for $\xi > \xi_2$.
We remove the index $i$ from $\mathcal{B}$ and replace it
by the index $j$. Since $i$ was chosen to be the earliest violated inequality,
after this dual simplex step the new basis defines a feasible and optimal
point for sufficiently small $\xi > \xi_2$.

From the viewpoint of the game, we can distinguish the following subcases
corresponding to the set of $m+n$ active inequalities.
\begin{enumerate}
\item \emph{$m+1$ inequalities for $x$ and $n-1$
inequalities for $y$ are binding.}

Then $i$ and $j$ refer to indices of inequalities for the polyhedron $P$.
If $i \in \{1, \ldots, m\}$ then one of the unplayed pure strategies of the first player
  is now effectively played.
If $i \in \{m+1, \ldots, m+n\}$ then one of the previous best pure responses of the second 
  player becomes a suboptimal response.

If $j \in \{1, \ldots, m\}$ then one of the played pure strategies of the
first player becomes unplayed.
If $j \in \{m+1, \ldots, m+n\}$ then one of the previous suboptimal pure responses 
of the second player becomes a best response.

\item \emph{$m$ inequalities for $x$ and $n$ inequalities
for $y$ are binding.}

Then $i$ and $j$ refer to indices of inequalities for the polyhedron $Q$.
If $i \in \{1, \ldots, m\}$ then one of the previous best pure responses of the first player
  becomes a suboptimal response.
If $i \in \{m+1, \ldots, m+n\}$ then one of the unplayed pure strategies of the second player
  is now effectively played.

If $j \in \{1, \ldots, m\}$ then one of the previous suboptimal pure responses
of the first player becomes a best response.
If $j \in \{m+1, \ldots, m+n\}$ then one of the played pure strategies of the second player 
  becomes unplayed.
\end{enumerate}

\medskip

\noindent
{\bf Case $\xi_2 = \beta_2$.} Then for the value $\xi = \xi_2$ there exists
an index $i \in \mathcal{B}$ such that the dual variable $y_i$ becomes zero.
For $\xi > \xi_2$ the current vector $z(\xi)$ is no longer optimal.
We remove the index $i$ from the basis and perform a simplex step moving
along an edge of the polyhedron. Since $i$ was chosen to be the earliest
violated optimality condition, this simplex step gives a new optimal 
basis $\mathcal{B'}$.

We conclude:

\begin{thm}
Let $(A,B)$ be a non-degenerate bimatrix game of rank~1. Then the parametric 
algorithm enumerates all Nash equilibria.
The running time of the algorithm is polynomial in the product $f_0(P) \cdot f_0(Q)$,
where $f_0(P)$ and $f_0(Q)$ denote the number of vertices of $P$ and $Q$ respectively.
\end{thm}

\begin{proof}
For those parameter values $\xi$ where the basis does not change,
the optimal basis is given by an $m$-element subset 
$I \subset \mathcal{I}$ corresponding to
a vertex of $P$ and by an $(n-1)$-element subset $J \subset \mathcal{J}$ that can
be extended to an $n$-element subset characterizing a vertex of $Q$.
\end{proof}

As mentioned in Section~\ref{se:prelim} note that in general not every vertex pair
of $P$ and $Q$ corresponds to a Nash equilibrium.

\subsection{Degenerate games\label{se:degenerate}}

If the game is degenerate then there are two issues. 
The polyhedra might not be simple
and the number of Nash equilibria can become infinite.
In order to resolve the first of these points, 
we have to cope with the same issues
as in the case of the simplex algorithm (in particular,
the issue of possible cycling when changing a basis).
However, with
the same techniques as for linear programming (such as symbolic perturbation,
lexicographic ordering), these situations can be resolved.

In order to extend the algorithm to degenerate games 
with an infinite number of equilibria
as well, by Section~\ref{se:earlierwork} the extreme equilibria are
sufficient to determine all maximal Nash subsets.
Indeed, our method can be modified to find all extreme equilibria
even in degenerate cases.

\subsection{Example\label{se:example}}

We consider the rank 1-game from~\eqref{eq:rank1constr} for $d=2$:
\[
  A \ = \ \left( \begin{array}{cc}
       2 & 7 \\
       1 & 8
\end{array} \right) \, , \qquad
  B \ = \ \left( \begin{array}{cc}
       2 & 1 \\
       7 & 8
      \end{array} \right) \, ;
\]
i.e., $A+B = b c^T$ with $b = (2,4)^T$, $c = (2,4)^T$.
We have $\min_{y \in \mathcal{S}_2} c^T y = 2$ and
$\max_{y \in \mathcal{S}_2} c^T y = 4$.
For the value of $\xi = 2$, the inequalities with indices in
$\{2,3,5,8\}$ are binding.
\[
\begin{array}{|l|r|l|} \hline
  && \\ [-2ex]
  \xi = c^T y & \text{objective} & \text{binding} \\ \hline \hline
  2 & 0 & \{2,3,5,8\} \\
  \xi \in (2, \frac{5}{2}) & < 0 & \{2,3,5\} \\
  \frac{5}{2} & - \frac{1}{4} & \{2,3,4,5\}  \\
  \xi \in (\frac{5}{2}, 3) & < 0 & \{3,4,5\} \\
  3 & 0 & \{3,4,5,6\} \\
  \xi \in (3,\frac{7}{2}) & < 0 & \{3,4,6\} \\
  \frac{7}{2} & - \frac{1}{4} & \{1,3,4,6\} \\
  \xi \in (\frac{7}{2},4) & < 0 & \{1,4,6\} \\
  4 & 0 & \{1,4,6,7\} \\ \hline
\end{array}
\]

For $\xi = \frac{5}{2}$, in the optimal situation we obtain uniquely $y=(\frac{3}{4},\frac{1}{4})$ and
$\pi_1 = \frac{13}{4}$. The optimal values for $x$ and $\pi_2$ are not unique; by 
substituting $x_2 = 1-x_1$ we can analyze the situation locally around the parameter
value $\xi = \frac{5}{2}$ in the $(x_1,\pi_2)$-plane
(see Figure~\ref{fi:notunique}; but note that the $x$ and $y$-axis are scaled differently).
The induced optimization problem is
\[
  \begin{array}{rcl@{\qquad\qquad}c}
  \multicolumn{3}{l}{\max~ (-2x_1+4) \xi - \pi_2 - \frac{13}{4}} \\
  0 & \le & x_1 \ \le \ 1 \, , & (1,2) \\
  -5 x_1 + 7 & \le & \pi_2 \, , & (3) \\
  -7 x_1 + 8 & \le & \pi_2 \, . & (4) \\
  \end{array}
\]
For $\xi = \frac{5}{2}$ the objective function is $-5 x_1 - \pi_2 + \frac{27}{4}$, so that
both points $p_1:=(1,2)^T$ and $p_2:=(\frac{1}{2},\frac{9}{2})^T$ (as well
as all convex combinations) are optimal. The first one comes from the
basis $\{2,3,5\}$ and the second one from the basis $\{3,4,5\}$.
For some sufficiently small~$\varepsilon$, in the case of $\xi = \frac{5}{2} - \varepsilon$,
the first of these bases is optimal, and in the case of $\xi = \frac{5}{2} + \varepsilon$
the second of these bases is optimal.

\ifpictures
\begin{figure}[ht]
\[
  \begin{array}{c@{\qquad \qquad}c}
    \includegraphics[scale=1]{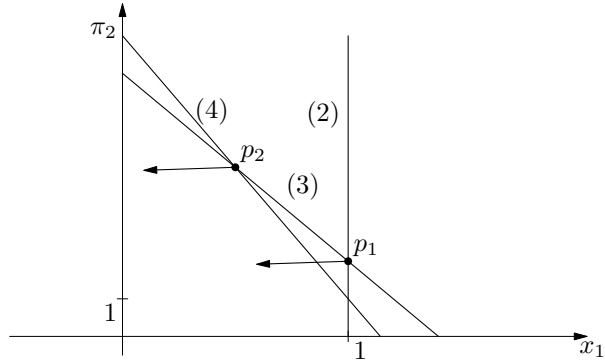}
  \end{array}
\]

\caption{The situation for $\xi = \frac{5}{2}$. Since the $x_1$- and 
$\pi_2$-axes
are scaled differently, the outer
normal vectors to the lines do not look orthogonal in
the figure.}
\label{fi:notunique}
\end{figure}
\fi

\section{Conclusion and outlook}

We have presented an enumeration algorithm for the Nash
equilibria of non-degenerate rank 1-games which is adapted to the
rank 1-structure. Our running time estimate was based on bounding the
number of vertices of the polyhedra involved. An open question is in
how far one can improve the running time analysis of the algorithm.

A widely open question is to develop enumeration algorithms for
games of rank $k$ (for fixed $k \in \N$) exploiting the
low-rank structure.

\subsection*{Acknowledgments.} Thanks to the reviewers for
very helpful comments and corrections.

\providecommand{\bysame}{\leavevmode\hbox to3em{\hrulefill}\thinspace}
\providecommand{\MR}{\relax\ifhmode\unskip\space\fi MR }
\providecommand{\MRhref}[2]{%
  \href{http://www.ams.org/mathscinet-getitem?mr=#1}{#2}
}
\providecommand{\href}[2]{#2}


\begin{thebibliography}{10}

\bibitem{aggarwal-73}
V.~Aggarwal. On the generation of all equilibrium points
for bimatrix games through the Lemke-Howson algorithm.
\emph{Math.\ Program.} 4:233-234, 1973.

\bibitem{ahjs-01}
C.~Audet, P.~Hansen, B.~Jaumard, and G.~Savard.
Enumeration of all extreme equilibria of bimatrix games.
\emph{SIAM J.\ Scientific Comput.} 23:323--338, 2001.

\bibitem{chen-deng-2005}
X.~Chen and X.~Deng. 
Settling the complexity of 2-player Nash equilibrium.
\emph{Proc.\ Foundations of Computer Science} (Berkeley, CA),
261--272, 2006.

\bibitem{cdt-2006}
X.~Chen, X.~Deng, and S.-H. Teng. 
Computing Nash equilibria: Approximation and smoothed complexity.
\emph{Proc.\ Foundations of Computer Science} (Berkeley, CA)
603--612, 2006.

\bibitem{conitzer-sandholm-2003}
V.~Conitzer and T.~Sandholm. 
Complexity results about Nash equilibria.
\emph{Proc. International Joint Conference on Artificial Intelligence}
(Acapulco, Mexico), 765-771, 2003.
 
\bibitem{dantzig-b51}
G.B.~Dantzig.
\emph{Linear Programming and Extensions.}
Princeton Univ.\ Press, Princeton, NJ, 1963.

\bibitem{isaacson-millham-80}
K.~Isaacson and C.B.~Millham.
On a class of Nash-solvable bimatrix games and some
related Nash subsets.
\emph{Naval.\ Res.\ Logist.\ Quarterly}
23:311--319, 1980.

\bibitem{jansen-81}
M.J.M.~Jansen. Maximal Nash subsets for bimatrix games.
\emph{Naval Research Logistics Quarterly} 28:147--152, 1981.

\bibitem{kannan-theobald-soda-2007}
R.~Kannan and T.~Theobald.
Games of fixed rank: A hierarchy of bimatrix games.
In \emph{Proc.\ Symposium on Discrete Algorithms} (New Orleans, LA), 2007.

\bibitem{konno-kuno-92}
H.~Konno and T.~Kuno.
Linear multiplicative programming.
\emph{Math.\ Program.} 56:51--64, 1992.

\bibitem{ktt-97}
H.~Konno, P.T.~Thach, and H.~Tuy.
\emph{Optimization on Low Rank Nonconvex Structures.}
Kluwer, Dordrecht, 1997.

\bibitem{lemke-howson-64} 
C.E.~Lemke and J.T.~Howson.
Equilibrium points of bimatrix games.
\emph{J.\ Soc.\ Indust.\ Appl.\ Math.} 12:413--423, 1964.

\bibitem{mangasarian-64}
O.L.~Mangasarian.
Equilibrium points of bimatrix games.
\emph{J.\ Soc.\ Indust.\ Appl.\ Math.} 12:778--780, 1964.

\bibitem{mangasarian-stone-64}
O.L.~Mangasarian and H.~Stone.
Two-person nonzero-sum games and quadratic programming.
\emph{J.\ Math.\ Anal.\ Appl. 9:348--355, 1964.}

\bibitem{millham-74}
C.B.~Millham. On Nash subsets of bimatrix games.
\emph{Naval Res.\ Logist.} 74:307--317, 1974.

\bibitem{mills-60}
H.~Mills.
Equilibrium points in finite games.
\emph{J.~Soc.\ Indust.\ Appl.\ Math.} 8:397--402, 1960.

\bibitem{nash-pams-50}
J.~Nash. Equilibrium points in $n$-person games.
\emph{Proc.\ Amer.\ Math.\ Soc.} 36:48--49, 1950.

\bibitem{mukhamediev-78}
B.M.~Mukhamediev.
The solution of bilinear programming problems and finding the
equilibrium situations in bimatrix games.
\emph{Comput.\ Math.\ Math.\ Phys.} 18:60--66, 1978.

\bibitem{nash-annals-51}
J.~Nash. Non-cooperative games. \emph{Ann.\ Math.} 54:286--295,
1951.

\bibitem{papadimitriou-2001}
C.H.\ Papadimitriou.
Algorithms, games and the Internet.
In \emph{Proc.\ 33rd ACM Symposium on Theory of Computing}
  (Chersonissos, Kreta), 749--753, 2001.

\bibitem{shapley-74}
L.S.~Shapley.
A note on the Lemke-Howson algorithm.
\emph{Math.\ Program.\ Study} 1:175--189, 1974.

\bibitem{stengel-99}
B.~von Stengel.
New maximal numbers of equilibria in bimatrix games.
\emph{Discrete Comput.\ Geom.} 21:557--568, 1999.

\bibitem{stengel-handbook-2002}
B.~von Stengel. Computing equilibria for two-person games.
In R.J.~Aumann, S.~Hart (Hrsg.),
\emph{Handbook of Game Theory}, North-Holland, Amsterdam, 2002.

\bibitem{theobald-semesterberichte-2005}
T.~Theobald.
Geometrie und Kombinatorik von Nash-Gleichgewichten.
\emph{Math.\ Semesterber.} 52:221--239, 2005.

\bibitem{vorobev-58}
N.N.~Vorob'ev. Equilibrium points in bimatrix games.
\emph{Theory Prob.\ Appl.} 3:297--309, 1958.

\bibitem{winkels-79}
H.M.~Winkels. An algorithm to determine all equilibrium points of a bimatrix game.
In O.~Moeschlin and D.~Pallaschke (eds.), \emph{Game Theory and Related Topics},
North-Holland, Amsterdam, 137--148, 1979.

\end{thebibliography}

\end{document}